\documentclass[aps,pra,showpacs,groupedaddress,superscriptaddress,twocolumn,longbibliography]{revtex4-2}

\usepackage{dsfont}
\usepackage{float}
\usepackage{xcolor}
\usepackage{physics}
\usepackage[caption=false]{subfig}

\usepackage[normalem]{ulem}
\usepackage{hyperref}
\hypersetup{
	breaklinks=true,
	colorlinks=true,
	allcolors=blue,
}

\usepackage{txfonts}
\usepackage{amsmath}
\usepackage{amssymb}
\usepackage{graphicx}% Include figure files
\usepackage{dcolumn}% Align table columns on decimal point
\usepackage{bm}% bold math
\usepackage{array}
\usepackage{multirow}

\begin{document}
\title{\emph{One knob to tune them all:} \\ Phase-controlled photon statistics and linewidth  in  partially pumped atomic ensembles}

\author{Oksana Chelpanova\href{https://orcid.org/0000-0002-1679-1359}{\includegraphics[height=1.7ex]{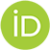}}}
\email{ochelpan@buffalo.edu}

\affiliation{Department of Physics, The State University of New York at Buﬀalo, Buﬀalo, New York 14260, USA}%Lines break automatically or can be forced with \\
  
\author{Martino Stefanini\href{https://orcid.org/0000-0002-8095-9603}{\includegraphics[height=1.7ex]{orcid-logo.png}}}%
\email{martino.stefanini@ictp-saifr.org}
\affiliation{Department of Physics, The State University of New York at Buﬀalo, Buﬀalo, New York 14260, USA} 
\affiliation{%
ICTP South American Institute for Fundamental Research
Instituto de Física Teórica, UNESP - Univ. Estadual Paulista
Rua Dr. Bento Teobaldo Ferraz 271, 01140-070, São Paulo, SP, Brazil
}%

\author{Dusan Sarenac\href{https://orcid.org/0000-0001-8575-3367}{\includegraphics[height=1.7ex]{orcid-logo.png}}}%
\affiliation{Department of Physics, The State University of New York at Buﬀalo, Buﬀalo, New York 14260, USA}

\author{Tim Thomay\href{https://orcid.org/0000-0003-2271-6803}{\includegraphics[height=1.7ex]{orcid-logo.png}}}%
\affiliation{Department of Physics, The State University of New York at Buﬀalo, Buﬀalo, New York 14260, USA}

\author{Jamir Marino\href{https://orcid.org/0000-0003-2585-2886}{\includegraphics[height=1.7ex]{orcid-logo.png}}}
\affiliation{Department of Physics, The State University of New York at Buﬀalo, Buﬀalo, New York 14260, USA}

\begin{abstract}
We study a minimal model of {collective} light emission from an {incoherently driven} ensemble of atoms {where incoherent drive is applied to only a subset of the atoms}  and show that both the linewidth and the photon statistics can be controlled within a single framework. 
In this setting, collective dissipation induces correlations between the pumped and unpumped parts of the system, leading to interference between their emission contributions. By introducing a relative phase between these contributions and tuning the pump rate, we demonstrate that the properties of the emitted light can be varied over a broad range. In particular, the linewidth can be made either independent of system size or scale extensively with it, while the photon statistics can be tuned from antibunched or quantum to bunched. 
{We further show that the role of the relative phase in controlling the interference can alternatively be played by the coherent interaction.}
By tuning the interaction strength together with the pump rate, one can access the same regimes as in the dissipation-only model. In addition, coherent interactions stabilize regimes of coherent emission with narrow linewidth, reminiscent of superradiant lasing. Our results illustrate how interference in partially driven collective systems provides a flexible mechanism for tailoring both spectral and statistical properties of light.
\end{abstract}

\maketitle

\section{Introduction}

\begin{figure*}
    \centering
    \includegraphics[width=1\linewidth]{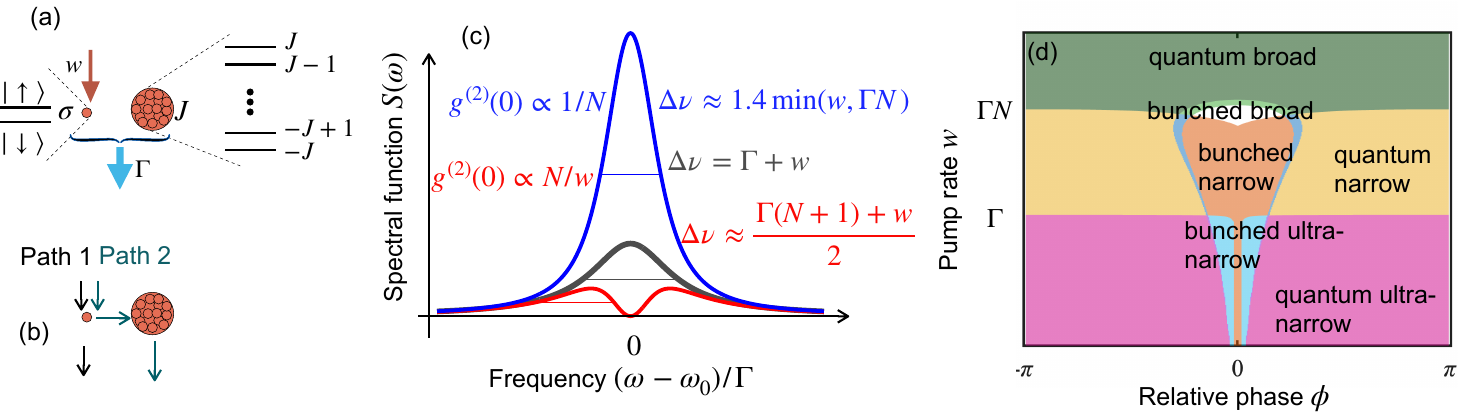}

    \caption{
(a) Schematic of the setup. A single spin $\sigma$ is incoherently pumped at rate $w$ and decays collectively together with the collective spin $J$. 
(b) Illustration of the two pathways in Liouvillian processes by which excitations are transferred from the pump to the environment: path 1 corresponds to excitation and subsequent emission by the pumped atom, while path 2 corresponds to transfer of the excitation to the unpumped spin $J$, followed by emission into the environment. 
(c) Spectral function of a single atom driven to the steady state by incoherent pumping $w$ and loss $\Gamma$ (black line), the spectral function for the setup in panel (a) (red line), and the spectral function for the same setup when the relative phase $\phi=\pi$ is imprinted in the dissipator (blue line). 
(d) Diagram summarizing regimes of the emitted light in terms of linewidth and light statistics, achievable by tuning the pump rate and the relative phase $\phi$. The full color palette is labeled in Fig.~\ref{fig:int}.
} 
    \label{fig:schematics}
\end{figure*}

{The statistical and spectral properties of light emitted by quantum optical sources determine their range of applications. For instance, sources of antibunched, or quantum, light, often realized as single-photon sources, are used in quantum information protocols such as quantum communication~\cite{gao2023atomically,lounis2005single,bennett2014quantum,Kirton2025collectiveenhancementphotonblockade,larsen2025chipscaleatomicbeamsource,PhysRevLett.109.163601}. Laser light—coherent in phase yet exhibiting essentially uncorrelated intensity fluctuations—is widely used in applications ranging from GPS to precision measurements and atomic clocks~\cite{Haroche_laser}. Intensity-correlated, or bunched, light likewise plays a central role in Hanbury Brown and Twiss–type experiments~\cite{10.1093/mnras/stac489,bromberg2010hanbury}. The linewidth and temporal coherence of light, on the other hand, become crucial when considering precision measurements. Applications such as atomic clocks and gravitational-wave detectors require long-lived phase coherence, or equivalently a narrow linewidth, to achieve high measurement precision~\cite{Kolkowitz,Kolkowitz2,bothwell2022resolving,optical_clock,Holland_SR_laser,kessler2012sub,jager2021regular,reilly2025fullycollectivesuperradiantlasing,Thompson_laser,PhysRevLett.130.223402,PhysRevX.8.021036,Abbott_2009}. At the same time, broad-linewidth light can also be advantageous, for example in absorption spectroscopy where a wide spectral coverage is desirable~\cite{coddington2016dual}. In this context, the ability to control both the photon statistics and the linewidth of emitted light within a single platform could enable the realization of a versatile, multipurpose light source.}

Various approaches to control  photon statistics have been explored in recent years. These include realizing photon blockade to generate quantum light~\cite{Kirton2025collectiveenhancementphotonblockade,PhysRevLett.134.183601}, or engineering burst-like emission to produce strongly bunched light~\cite{superbunched_Kim,doi:10.1126/sciadv.aav4986,Solomon}.
A conceptually different route relies on interference effects: when multiple emission channels are present, the output field is given by their coherent superposition, enabling control over photon correlations~\cite{von_Zanthier_optica,vonZanthier_phase_control,jing2026manybodyamplifiednonclassicalphoton}. For example, in Ref.~\cite{vonZanthier_phase_control}, competition between superradiant and subradiant modes allows one to tune photon statistics from quantum to bunched by modifying the position of a single atom inside a cavity.

Parallel progress has been made in controlling the emission linewidth. {In single-emitter systems, the emission linewidth of semiconductor quantum dots has been shown to narrow under the addition of weak above-band excitation to a resonant drive~\cite{Gazzano:18}. In many-body systems,} superradiant lasers operating in the bad-cavity regime provide a route to narrow-linewidth emission that is largely insensitive to cavity-length fluctuations, as the coherence is stored in the atomic degrees of freedom~\cite{Holland_SR_laser,Thompson_laser,zhang2024extremely}.
Additional improvements have been achieved using filter cavities~\cite{filter_cavities,sbbk-xdvs} or multilevel schemes~\cite{reilly2025fullycollectivesuperradiantlasing}. More recently, attention has turned to the role of inhomogeneities, such as spatially nonuniform incoherent driving. It has been shown that the presence of unpumped emitters can modify the linewidth, leading to broadening in cavity emission~\cite{Bruder_laser} or narrowing for light emitted in free space~\cite{Ritsch_partial_laser,Ritsch_anthena}. These observations suggest that inhomogeneous driving can qualitatively alter the emission properties, although a unifying physical picture is still developing.

In this work, we explore a minimal setting in which both the linewidth and the photon statistics can be tuned within a single framework by exploiting interference between emission pathways. Our approach is based on a simple model inspired by partially pumped systems~\cite{Ritsch_partial_laser,Bruder_laser}, in which an incoherent drive repumps only a subset of atoms, while all atoms experience collective decay, see Fig.\ref{fig:schematics}(a). In this situation, emission can proceed either directly from the pumped subsystem, which is subject to both pump and loss noise, or indirectly via the unpumped subsystem, which experiences only loss, cf. Fig.\ref{fig:schematics}(b). These two contributions then interfere. The interference originates from the coherence between the subsystems induced by collective dissipation and persists in the steady state.

To access and control this interference, we introduce a relative phase $\phi$ between the emission pathways, which can be interpreted as a parameter of {either of the detected field or of the system}. Within this simplified model, we show that tuning $\phi$ modifies both spectral and statistical observables. In particular, for $\phi=0$ {(fully symmetric decay channel)} the spectral function develops a dip at zero detuning and a two-peak structure due to destructive interference between the two contributions, see Fig.~\ref{fig:schematics}(c). In this regime, the linewidth is not set by the slowest decay rate {of the system}, but rather by a subleading mode. By tuning the phase toward $\phi=\pi$, the interference becomes constructive, leading to a crossover from broad to narrow emission.

We further show that this mechanism allows one to vary photon statistics and linewidth largely independently. Depending on the pump strength and phase, the system can emit antibunched (quantum), coherent, or bunched light, while the linewidth can remain narrow or scale extensively with system size, see Fig.~\ref{fig:schematics}(d). In particular, we identify regimes in which  quantum light has a narrow linewidth, as well as regimes of bunched emission with reduced spectral width. These features persist across system sizes and reflect the interplay between collective dissipation and inhomogeneous driving. The two-spin limit of our model captures the essential physics of single-emitter experiments with a secondary incoherent channel~\cite{Solomon,Gazzano:18}, suggesting that signatures of partial pumping may be accessible via photon-number-resolving detection~\cite{Thomay_multiphoton}.

Finally, we extend the analysis by including coherent interactions between the pumped and unpumped subsystems {and fixing $\phi=0$}. In this case, the interaction provides a natural phase reference, and tuning the interaction strength and pump rate allows one to access similar regimes as in the {purely} dissipative model. In addition, the coherent coupling stabilizes further dynamical behavior, including population inversion and emission with characteristics reminiscent of superradiant lasing. The resulting phase diagram reflects the combined influence of pumping and coherent interactions.

The paper is organized as follows. In Section~\ref{sec:model}, we introduce the toy model and the set of observables evaluated in the steady state. In Section~\ref{sec:dissip_only}, we analyze the interference between the two emission pathways, starting from the two-spin case and then presenting results for a many-body system. In Section~\ref{sec:int}, we extend the analysis to include coherent interactions, and Sec.~\ref{sec:discussion} contains conclusions and discussion. The Appendices provide details on the exact numerical solution, an analysis of a dissipative model supporting ultranarrow thermal light, a possible experimental implementation with independent control of the phase $\phi$, and additional technical details.

\section{Model\label{sec:model}}

We consider an ensemble of $N+1$ two-level atoms, modeled by pseudo-spin-$1/2$ operators $\sigma_i^\alpha$, $i=1,\,2,\dots,\,N+1$, obeying the $\mathrm{SU}(2)$ algebra, see Fig.~\ref{fig:schematics}(a). All atoms undergo collective decay at rate $\Gamma$, while one atom is additionally incoherently pumped at rate $w$.  
{Since the incoherent processes in this problem are either fully collective or local, we do not specify the spatial distribution of atoms. In Appendix~\ref{sec:exp}, we outline a possible cavity QED implementation, where uniform coupling to a leaky cavity mode induces collective decay, while a local laser field can be used to address a single atom and induce a local spin pump. Without loss of generality, we set the spin $\sigma_{N+1}\equiv \sigma$ to be incoherently pumped} 
and combine the remaining  $N= 2J$ atoms participating only in the decay process into a collective pseudo-spin operator $J^{\alpha}=\sum_{i=1}^N \sigma_i^{\alpha}$.

The dynamics of the system are governed by the Lindblad master equation~\cite{breuer2002theory,stefanini2025lindblad,fazio2025many}
\begin{equation}\label{eqref:model}
\dot \rho=\mathcal L\rho=
\frac{\Gamma}{2}\mathcal{D}\!\left[e^{\mathrm{i}\phi}\sigma^-+J^-\right]\,\rho
+\frac{w}{2}\mathcal{D}\!\left[\sigma^+\right]\,\rho ,
\end{equation}
where the dissipator is defined as
$\mathcal{D}[A]\,\rho=2 A\rho A^\dagger-\{A^\dagger A,\rho\}$.
Here, we  introduce a relative phase $\phi$ between the pumped and unpumped spins, where
for $\phi=0$, the loss term describes symmetric collective decay, which can be realized, for example, when all $N+1$ atoms are coupled to a single lossy cavity mode~\cite{Holland_SR_laser}.  
Importantly, most of the results presented in this paper rely on the ability to control the phase in the jump operator while keeping the measured observables insensitive to $\phi$, or vice versa.  
One possible route to realizing such a scenario experimentally is discussed in the Appendix~\ref{sec:exp}.

The model~\eqref{eqref:model} possesses a strongly conserved quantity, the {total $J$ angular momentum} {of the unpumped subsystem}, $\bm{J}^2=\sum_\alpha (J^\alpha)^2=J(J+1)$, which implies that its dynamics occur in independent manifolds spanned by the Dicke states, namely eigenstates of $\bm{J}^2$ and $J^z$~\cite{breuer2002theory}. We choose to study its behavior in the subspace with maximal $J=N/2$, which can be accessed, e.g., by initializing all atoms in their ground state.
In this case, the Hilbert space of the model scales as $4(N+1)^2$, which allows one to perform numerics for tens of spins without difficulty. Additionally, by restricting the number of excitations in spin $J$, the numerics can be extended further to larger system sizes. Further discussion of the exact solution is provided in the Appendices~\ref{sec:ED} and~\ref{sec:semiclassics}.

\subsection{Observables\label{sec:observables}}

To analyze the steady-state properties of the emitted light, we evaluate several steady-state observables~\cite{walls2008quantum}. The first observable is the total magnetization
$\langle S^z\rangle=\langle \sigma^z+J^z\rangle$,
which is related to the number of excitations in the system. We also evaluate $\langle S^+ S^-\rangle$, which is proportional to the emitted light intensity, where $S^{\pm}=J^{\pm}+\sigma^{\pm}$. 
Importantly, $S^{\pm}$ do not contain the phase $\phi$, which allows the interference effects discussed below to be observed and controlled. Alternatively, one may consider a phase-insensitive jump operator $L=\sigma^- + J^-$ and instead introduce the phase in the measured operators, $S^- = e^{-\mathrm i\phi}\sigma^- + J^-$, (one can switch between the two representations by a unitary rotation of the pumped spin). In this sense, the phase $\phi$ is not merely a gauge choice, but a physically relevant parameter, as we assume that it can be independently tuned in the observables and in the Liouvillian; see the further discussion in the Appendix~\ref{sec:exp}.

Another quantity we monitor is the equal-time second-order coherence~\cite{walls2008quantum,PhysRevX.7.041036}
\begin{equation}
  g^{(2)}(0)=\frac{\langle S^+ S^+ S^- S^-\rangle}{\langle S^+ S^-\rangle^2}.  
\end{equation}
Values of $g^{(2)}(0)<1$ correspond to antibunched light, in which photons are emitted one by one. Such a regime cannot be captured by classical theories of light and is therefore termed nonclassical, or quantum (Q), light. The case $g^{(2)}(0)=1$ indicates Poissonian (classical coherent) light~\footnote{{For finite system sizes, $g^{(2)}(0)$ is slightly larger than $1$ due to the $1/N$ corrections.}}, while $g^{(2)}(0)>1$ corresponds to bunched light. 
To further characterize the nature of the emitted light, in particular to distinguish between different types of bunched statistics (e.g., dominated by few-photon processes or exhibiting thermal-like behavior), we also evaluate higher-order coherences $g^{(k)}(0)={\langle (S^+)^k (S^-)^k\rangle}/{\langle S^+ S^-\rangle ^k}$ {up to $k=3$}. {The higher-order coherences $g^{(k)}(0)$  have become experimentally accessible through recent advances in photon-number-resolving detectors, including both transition-edge sensors and superconducting nanowire detectors, enabling engineering and classification of multiphoton states~\cite{Migdall,10.1063/5.0204340,magana2019multiphoton,Thomay_multiphoton}.}

The spectral properties of the emitted light are encoded in the spectral function
\begin{equation}\label{eq:spectral_f}
S(\omega)=2\Re \int dt\, e^{\mathrm{i}\omega t}\,
\langle S^+(t)S^-(0)\rangle,
\end{equation}
where expectation value $\langle\cdot\rangle$ is evaluated in steady state. From the spectral function, we extract the linewidth $\Delta \nu$ as the full width at half maximum (FWHM). In Fig.~\ref{fig:schematics}(c)
 we indicate the linewidth with thin solid lines. If the spectral function contains two symmetric peaks (red line), the linewidth is calculated as the FWHM of one of the peaks. {In order to distinguish between different emission regimes, we compare the linewidth to the characteristic decay rates of a single atom, $\Gamma$, and of collective decay for $N$ atoms, $\Gamma N$. If the linewidth $\Delta\nu$ does not scale with $N$ and is of order of $2\Gamma$~\footnote{Numerically, we find $\Delta\nu\lesssim 2.5\Gamma$ as a reasonable cutoff for the ultranarrow regime.}, 
 we refer to it as ultranarrow (UN). If the linewidth scales with $N$ or the pump rate $w$, but remains below $\Gamma N$, we classify it as narrow (N). If the linewidth exceeds $\Gamma N$, we refer to it as broad (B)}.
 
 We also evaluate how much the position of the peak of $S(\omega)$ deviates from zero detuning and denote this frequency shift as $\Delta \omega$. Here, we assume that both spins $\sigma$ and $J$ have the same level splitting $\omega_0$ and work in a rotating frame at this frequency. All frequencies and detunings are therefore defined with respect to $\omega_0$.

For the following discussion, it is useful to relate the shape of the spectral function to the spectrum of the Liouvillian.
Applying the quantum regression theorem~\cite{carmichael2013statistical} and the spectral decomposition of $\mathcal L$~\cite{Schiro2019spectral}, the two-time correlation function in Eq.~\eqref{eq:spectral_f} can be written as 
\begin{equation}
\begin{aligned}
\langle S^+(t)S^-(0)\rangle
&=\text{tr}\!\left(S^+e^{\mathcal L t}(S^-\rho_{ss})\right) \\
&=\sum_k e^{\lambda_k t}
\,\text{tr}(r_k S^+)
\,\text{tr}(l_k^\dagger S^- \rho_{ss}),
\end{aligned}
\end{equation}
where $\lambda_k$ are the eigenvalues of the Liouvillian $\mathcal L$ and $l_k$, $r_k$ are the corresponding left and right eigenvectors.

The spectral function then can be expressed as
\begin{equation}\label{eq:decomposition}
\begin{aligned}
S(\omega)
&=2\Re \sum_k
\frac{\text{tr}(r_k S^+)\text{tr}(l_k^\dagger S^-\rho_{ss})}
{-i(\omega+\Im\lambda_k)-\Re\lambda_k}\\
&=2\Re \sum_k
\frac{c_k}{-i(\omega+\Im\lambda_k)-\Re\lambda_k}.
\end{aligned}
\end{equation}
Here, the coefficients (or residues) $c_k$ depend on the steady-state density matrix as well as on the eigenvectors of $\mathcal L$. Since the dissipator \eqref{eqref:model} can always be brought to its $\phi=0$ form by a unitary transformation $\sigma^-\to\sigma^- e^{-i\phi}$, the Liouvillian eigenvalues $\lambda_k$ are insensitive to $\phi$, while the eigenvectors and therefore the weights $c_k$ change with $\phi$. Note, that the eigenmodes can be naturally ordered in terms of their real part---a smaller value of $\abs{\Re\lambda_k}$ implies that the mode persists for longer timescales. We will refer to the latter modes as ``slower''.
The above decomposition of $S(\omega)$ will be used later to provide a convenient way to interpret our findings in terms of the slowest-decaying eigenmodes of the Liouvillian.

We emphasize that the results presented in the following section rely on the ability to tune the phase inside the jump operator $J^-+e^{\mathrm{i}\phi}\sigma^-$ while measuring observables corresponding to the symmetric operators $S^{\pm}=J^{\pm}+\sigma^{\pm}$ (or vice versa). If no phase transformation is applied to the emitted light, the measured result corresponds to $\phi=0$. When coherent interaction is added to the Hamiltonian, however, it naturally imprints a reference phase, and no additional control over $\phi$ is required.
In the following section, we describe the effect of partial pumping under idealized conditions.

\section{Control of observables in dissipation-only partially pumped system\label{sec:dissip_only}}

The spin loss term in Eq.~\eqref{eqref:model} can be rewritten as
\begin{equation}\label{eq:Gamma_contributions}
\begin{aligned}
\mathcal{D}[e^{\mathrm i \phi}\sigma^-+J^-]\,\rho
&=2\sigma^-\rho\sigma^+-\{\sigma^+\sigma^-,\,\rho\}\\
&\quad+2J^-\rho J^+-\{J^+J^-,\,\rho\}\\
&\quad+2e^{-\mathrm{i}\phi}J^-\rho\sigma^+ -e^{-\mathrm{i}\phi}\{\sigma^+J^-,\,\rho\}\\
&\quad+2e^{~\mathrm{i}\phi}\sigma^-\rho J^+-e^{~\mathrm{i}\phi}\{J^+\sigma^-,\,\rho\}.
\end{aligned}
\end{equation}
The first line describes the decay of a single atom. The second line describes collective decay of the large spin $J$ (with an effective decay rate $\propto \Gamma N$; see also Dicke superradiance~\cite{gross1982superradiance,breuer2002theory}). The last two lines describe the dissipative coupling between the spins $\sigma$ and $J$, and these are precisely the terms that make the physics of this model interesting.

Without the last two lines, the result would coincide with that of a single atom: only the first atom radiates. In this case, the second line simply describes a trivial steady state for the spin $J$ with zero excitations, since this part of the ensemble is not pumped and therefore does not radiate. As a consequence, the total magnetization reads $\langle J^z+\sigma^z\rangle =-J+ (w-\Gamma)/2(w+\Gamma)$, the intensity satisfies $\langle S^+S^-\rangle=\langle \sigma^+\sigma^-\rangle=1/2+\langle\sigma^z\rangle$,
and all higher-order correlation functions vanish. The spectral function is Lorentzian, with linewidth $w+\Gamma$, which is determined solely by the properties of a single spin.

After including the third and fourth lines of Eq.~\eqref{eq:Gamma_contributions}, the dynamics become more interesting.
The spectrum now consists of a broad Lorentzian with a dip at zero detuning (see the red line in Fig.~\ref{fig:schematics}(c)). The linewidth of the emitted light is given by
$\Delta\nu=(\Gamma (N+1)+w)/2$,
which can easily become larger than the single-atom linewidth $\Gamma+w$ (see the black line in the same figure). 
Additional changes appear in the second-order coherence $g^{(2)}(0)$. Instead of vanishing, as in the single-atom case, it now scales as $N\Gamma/w$. As a result, the emitted light can be bunched, $g^{(2)}(0)>1$, coherent, $g^{(2)}(0)\to 1$, or quantum, $g^{(2)}(0)<1$, upon increasing the pump rate.  As we show below, these features originate from interference processes that occur in partially pumped systems and can be controlled by tuning the relative phase $\phi$~\footnote{As mentioned above, one can alternatively keep the jump operator in the form $\mathcal{D}[\sigma^-+J^-]$ but redefine the observables as $S^{\pm}=e^{\pm \mathrm{i}\phi}\sigma^{\pm}+J^{\pm}$, which effectively introduces the same phase shift in the emitted field.}.

\begin{figure*}
         \centering
         \includegraphics[width=\linewidth]{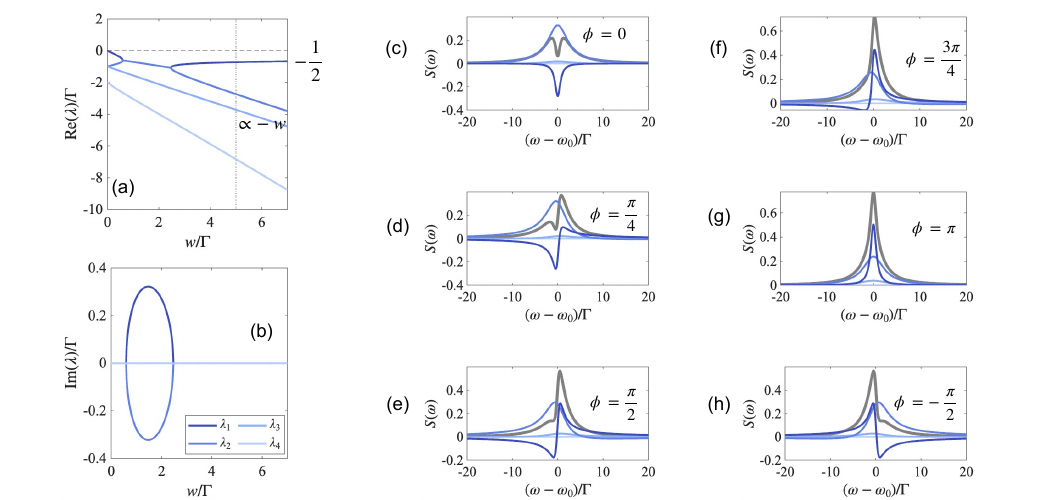}
         \caption{
         \textbf{Spectral decomposition of $S(\omega) $ for a system of two spins.}
         (a) Real and (b) imaginary parts of the Liouvillian eigenvalues for a system of $1+1$ spins. The real part of the slowest eigenvalue $\lambda_1$ saturates at the value $-\Gamma/2$, while the other eigenvalues decrease linearly with the pump rate $w$. 
(c–h) Contributions to the spectral function from different Liouvillian eigenmodes for $w=5\Gamma$, using Eq.~\eqref{eq:decomposition}. In the symmetric case (c), the eigenvalue $\lambda_1$ contributes to $S(\omega)$ with a negative sign, and the linewidth is determined by $|\Re \lambda_2|$. In the antisymmetric case (g), the interference is constructive and the linewidth is proportional to the real part of the slowest eigenvalue $|\Re \lambda_1|$. For intermediate values of $\phi$, the presence of the complex phase $e^{\mathrm{i}\phi}$ results in an antisymmetric shape of the spectral function.   
}
         \label{fig:two_spins}
 \end{figure*}

{In the following section, we explain the interference mechanism by analyzing the contributions of different Liouvillian eigenvalues to the spectral function for a system of two spins.}

\subsection{Two spins case\label{sec:2spins}}

To understand the origin of the dip at zero detuning, we first consider the simplest model consisting of two spins ($J=1/2$) and analyze the contribution of different Liouvillian eigenmodes to $S(\omega)$~\cite{Schiro2019spectral}. Note, that transient dynamics of collective decay of two emitters has been explored experimentally in Ref.~\cite{mlynek2014observation}. 
In Fig.~\ref{fig:two_spins}(a,b) we plot the eigenvalues of $\mathcal L$ as a function of the pump rate $w$ (note that the eigenvalues do not depend on the phase $\phi$). 
We show only those modes that contribute to the spectral function and have nonzero residues $c_k$ in Eq.~\eqref{eq:decomposition}. 
The real part of the slowest eigenvalue $\lambda_1$ starts from zero and then decreases, eventually saturating at $-\Gamma/2$ at strong pump. The real part of the second-slowest eigenvalue $\lambda_2$ (as well as the remaining eigenvalues $\lambda_{3,4}$) decreases linearly with the pump rate and does not saturate.
Naively, one might expect that $|\Re\lambda_1|$ determines the linewidth. However, as we show below, this is not always the case.

We fix $w=5\Gamma$ {(corresponding to well-separated $\lambda_k$)} and plot the contribution of different Liouvillian eigenmodes to the total spectral function for $\phi=0$, see Fig.~\ref{fig:two_spins}(c) \footnote{{This choice is made for simplicity, in order to keep all eigenvalues real. The case of complex eigenvalues and exceptional points is mentioned later in the text.}}. 
Contrary to the expectation, we find that the spectral width is  set by the second-slowest eigenvalue $\lambda_2$.
The narrow Lorentzian associated with $\lambda_1$ is still present in $S(\omega)$, but it appears with a negative residue, $c_1<0$. As a result, the contributions of the modes $\lambda_{1,2}$ interfere destructively, producing a dip at zero detuning in $S(\omega)$ (gray line). The position of this dip can be shifted by introducing a detuning between the pumped and unpumped spins; however, the linewidth of the spectral envelope remains determined by $|\mathrm{Re}\,\lambda_2|$.

The observation that the sign of residues of some eigenmodes $c_k$ is negative is not a generic feature, but rather a consequence of having collective spin loss in our toy model. We can organize the Hilbert space of the model into sectors with different numbers of excitations, coupled by the dynamics. We ignore the zero-excitation state given by $\ket{\downarrow \downarrow}$ since it does not contribute to the spectral function.  The one-excitation sector is spanned by the basis states $\ket{\uparrow \downarrow}$ and $\ket{\downarrow\uparrow}$. Without the pump, the steady state of the system in the single-excitation subspace is a singlet state, $\ket{\psi}\propto\ket{\uparrow\downarrow}-\ket{\downarrow\uparrow}$, for which the corresponding part of the density matrix $|\psi\rangle\langle\psi|$ has negative coherences. When the pump is added, it dresses the singlet state, but because of the collective nature of the loss, the coherences remain negative. As such, when calculating the contribution to the spectral function from cross terms $\langle \sigma^+(t)J^-(0)\rangle$ and $\langle J^+(t)\sigma^-(0)\rangle$ the corresponding contributions will be negative \footnote{This is essentially a continuity argument: $\expval{J^+(t=0)\sigma^-(0)}<0$ and $\langle \sigma^+(t=0)J^-(0)\rangle<0$, and thus remain negative (in real part) in a neighborhood of $t=0$. Since the decay at later times is exponential, the spectral function is mainly sensitive to the small-time behavior, especially close to the dip at $\omega=0$.}, in contrast to the positive contributions from $\expval{J^+(t)J^-(0)}$ and $\expval{\sigma^+(t)\sigma^-(0)}$. One can also deduce that the singlet-like state results in a lower intensity $\langle S^+ S^-\rangle$, which allows for $g^{(2)}(0)>1$ at weak pump.
This description is simplified, since the signs of elements of left and right eigenvectors components also bring their own corrections to $c_k$ [cf. the first line of Eq.~\eqref{eq:decomposition}], but overall intuition remains correct: cross terms give negative contribution to the spectral function~\footnote{
The unnormalized steady-state density matrix for the two-spin model reads
\begin{equation*}\label{eq:2spin_ss}
\rho_{ss}\propto v\left[\begin{array}{cccc}
\frac{w}{2\Gamma} & 0 & 0 & 0\\
0 & 1 & -\frac{w+2\Gamma}{2\Gamma}e^{\mathrm{i}\phi} & 0\\
0 & -\frac{w+2\Gamma}{2\Gamma}e^{-\mathrm{i}\phi} & \frac{6w\Gamma+w^{2}+2\Gamma^{2}}{2\Gamma^{2}} & 0\\
0 & 0 & 0 & \frac{w+4\Gamma}{2\Gamma}
\end{array}\right] v^T\end{equation*}
where  $v=\left[\begin{array}{cccc}
\ket{\uparrow \uparrow}, & \ket{\downarrow \uparrow}, & \ket{\uparrow \downarrow} , & \ket{\downarrow \downarrow}\end{array}\right]$ is the basis vector. Here, for example, $\ket{\uparrow \downarrow}$ denotes the state in which the pumped atom is excited and the unpumped atom is in the ground state. The sign of the coherences can be tuned via the phase $\phi$.
}.

Our numerics suggest that the slowest mode $\lambda_1$ is associated with the unpumped spin, since the latter is subjected only to noise originating from the loss channel, whereas the pumped spin $\sigma$ is additionally affected by noise from the pump and is therefore mainly responsible for the faster eigenmode $\lambda_2$. The slowest mode contributes predominantly to the correlation functions $\langle J^+(t)J^-(0)\rangle$ and $\langle J^+(t)\sigma^-(0)\rangle$, the latter being negative due to the antisymmetric, singlet-like structure of the steady state, as mentioned above. In contrast, $\lambda_2$ is mainly responsible for the contribution from $\langle \sigma^+(t)\sigma^-(0)\rangle$ and $\langle \sigma^+(t)J^-(0)\rangle$. The latter is negative, but an order of magnitude smaller than the others. As a result, the spectral function is dominated by a broad Lorentzian associated with emission from the $\sigma$ spin, while the slowest mode produces a narrow contribution of negative sign, mainly through the interference term $\langle J^+(t)\sigma^-(0)\rangle$, giving rise to the dip at zero detuning and making the linewidth effectively determined by $|\Re\lambda_2|$ rather than by the slowest decay rate.

\begin{figure*}
\centering
    \includegraphics[width=1\linewidth]{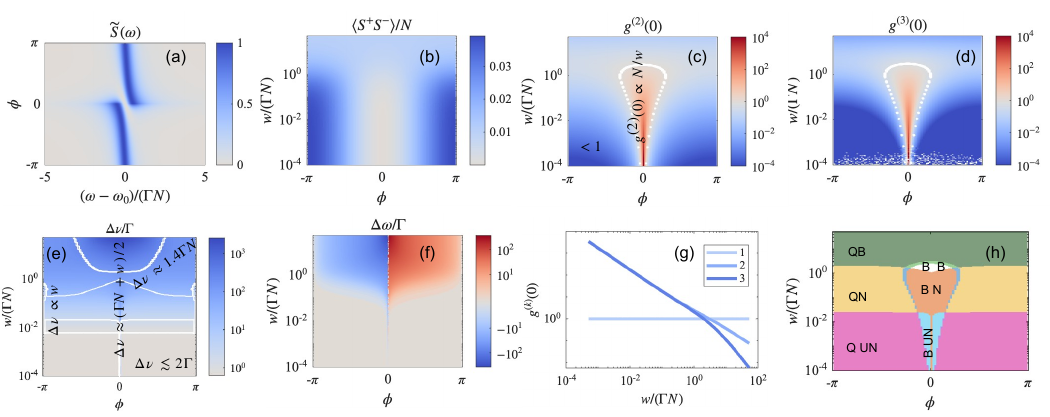}
    \caption{\textbf{Properties of the light emitted by an ensemble of $101$ atoms.} 
    (a) Normalized spectral function $\tilde S(\omega)=S(\omega)/\max(S(\omega))$ for different values of the phase $\phi$, with the pump rate fixed at $w=\Gamma N$.  
(b) Intensity, (c) second- and (d) third-order coherence, (e) linewidth, and (f) position of the largest peak (line shift) as functions of the phase $\phi$ and pump rate $w$. For $\phi$ close to $\pi$ the emitted light is always quantum and $g^{(2)}(0)\propto 1/N$, while for $\phi\approx 0$ the photon statistics can be tuned by varying $w$, with $g^{(2)}(0)\propto N\Gamma/w$. 
(g) Higher-order coherences as a function of $w$ for different $k$ and $\phi=0$. 
(h) Schematic of the accessible regimes summarizing the scaling of $g^{(2)}(0)$ and $\Delta\nu$. By tuning $w$ and $\phi$, the emitted light can be bunched ($g^{(2)}(0)>1$) with ultranarrow linewidth $\Delta\nu<2\Gamma$ (B UN), bunched with narrow linewidth $\Delta\nu<\Gamma N$ (B N), bunched with broad linewidth $\Delta\nu>\Gamma N$ (B B), or quantum ($g^{(2)}(0)<1$) with ultranarrow (Q UN), narrow (Q N), or broad (Q B) linewidth.}
 \label{fig:phase_plot}
\end{figure*}

Figures~\ref{fig:two_spins}(d–h) show that the situation changes significantly as the phase $\phi$ is varied. The eigenvalues $\lambda_k$ remain unchanged, but the eigenvectors $l_k$ and $r_k$ inherit the phase $\phi$, and the coefficients $c_k$ therefore become phase dependent.  
A complex $c_k$ contributes to the spectral function as $S_k(\omega)\propto \Re(c_k)\,|\Re(\lambda_k)|-\Im(c_k)(\omega+\Im(\lambda_k)),$
so that the second term produces an asymmetric spectral profile when it is nonzero. When $\phi=\pi$, the coefficient $c_1$ becomes positive and the interference becomes constructive, see Fig.~\ref{fig:two_spins}(g). 
The linewidth in this case is proportional to the slowest eigenvalue, $\Delta \nu\propto |\Re \lambda_1|$. Note, that for $\phi=\pi,$ the dark state at zero pump is a triplet state $\ket{\psi}\propto \ket{\uparrow\downarrow}+\ket{\downarrow\uparrow}$ and coherences give a positive contribution to the spectral function. Positive coherences also result in higher intensity $\langle S^+ S^-\rangle$ and, as a consequence, the light displays quantum (anti-bunched) statistics for all $w$. 
Overall, all coherence-dependent observables depend on the relative phase $\phi$, providing an additional mechanism for controlling the properties of the emitted light.

The plots in Fig.~\ref{fig:two_spins}(a,b) show two exceptional points at $w_1\approx 0.6 \,\Gamma$ and $w_2\approx 2.5\, \Gamma$, at which the first two eigenmodes of the Liouvillian coalesce~\cite{Ashida02072020}. For $w_1<w<w_2$, the two slowest eigenvalues share the same real part, while they acquire complex-conjugate imaginary parts. Despite this discontinuous spectral behavior, the observables we have analyzed behave in a continuous fashion across the exceptional points. 
If the eigenvalues $\lambda_{1,2}$ form a complex-conjugate pair, the qualitative result remains unchanged, although the individual contributions to the spectral function become asymmetric. The system nevertheless displays the same kind of interference.

\subsection{Many-body case}

{We now turn to larger system sizes and examine the manifestation of the interference mechanism identified in the two-spin case. As an illustration, we plot the normalized spectral function $\tilde S(\omega)=S(\omega)/\max(S(\omega))$ for a system with $N=100$ spins, evaluated for different values of the relative phase $\phi$ at a fixed pump rate $w=\Gamma N$, see  Fig.~\ref{fig:phase_plot}(a). The spectral function exhibits a transition from a double-peak to a single-peak structure as $\phi$ approaches $\pi$. }

We further study the dependence of the observables introduced in Sec.~\ref{sec:observables} for a system with $N=100$ as functions of the phase $\phi$ and pump rate $w$. The magnetization $\langle S^z\rangle$ does not depend on $\phi$ and exhibits only a weak dependence on the pump rate, varying from $-J-1/2$ to $-J+1/2$ as $w$ increases. We therefore focus on observables that depend on the off-diagonal elements of the density matrix: since these coherences depend on $\phi$, the corresponding observables can be tuned by adjusting the relative phase. 

Figures~\ref{fig:phase_plot}(b–f) show the intensity $\langle S^+ S^-\rangle$, the second- and third-order coherence at zero time delay, $g^{(2)}(0)$ and $g^{(3)}(0)$, the linewidth $\Delta\nu$, and the peak position $\Delta\omega$ as functions of $\phi$ and $w$. The intensity of the emitted light in panel (b) is approximately four times larger in the antisymmetric case $\phi=\pi$ than in the symmetric case $\phi=0$, although it does not scale with system size and therefore remains relatively small. {We explain variation of intensity with $\phi$ by the shape of one excitation sector of the density matrix: for $\phi=0$ the non-diagonal elements of the density matrix are negative, so the state is {``singlet-like''}, which results in lower intensity comparing to $\phi=\pi$ case, where the  non-diagonal elements of the $\rho_{ss} $ are positive and the state is {``triplet-like''}}.

The second-order coherence $g^{(2)}(0)$ behaves differently in the symmetric and antisymmetric cases. For $\phi=0$, we find $g^{(2)}(0)\approx1.7\, N\Gamma/w$, so that the emitted light can transition from bunched ($g^{(2)}(0)>1$) to quantum ($g^{(2)}(0)<1$) as the pump rate increases beyond $w\approx1.7\,\Gamma N$. However, the bunched regime is dominated by two-photon processes, as higher-order coherences $g^{(k)}(0)$ do not display fast growth with $k$, as in thermal light, cf. panels (d,g). 
We emphasize that this distinction is invisible at the level of $g^{(2)}(0)$ alone: bursty emission of photon pairs and thermal (chaotic) light can produce indistinguishable $g^{(2)}(0)$ values at appropriate pump rates.
The two processes diverge only at higher orders, where thermal light displays $g^{(k)}(0)\propto k!$ growth while the interference-induced bunching saturates.
This makes higher-order coherence measurements ($k \geq 3$) the natural discriminator between the mechanism described here and a thermal-bath origin of the bunching.
In the bunched regime, the destructive interference suppresses the intensity but keeps the intensity correlation function finite, which results in the $g^{(2)}(0)>2$~\cite{5ts6-nlrl}.
For the case $\phi=\pi$, the emitted light is always quantum, with a maximum value $g^{(2)}(0)\approx 0.5$ occurring near $w\propto\Gamma N$. Importantly, $g^{(2)}(0)$ decays as $1/N$, suppressing the probability of emitting two or more photons simultaneously as $N$ is increased. 

Although both symmetric and antisymmetric configurations allow the generation of quantum light, the linewidth behaves very differently in the two cases, see panel (e). In the $\phi\to 0$ configuration, the linewidth increases with both pump rate and system size, $\Delta\nu=({\Gamma(N+1)+w})/{2}$,
so quantum light is obtained only at the cost of a broad spectral function. In contrast, in the $\phi\to\pi$ case, the linewidth is controlled by $|\Re\lambda_1|\propto\min(w,\Gamma N)$ (with an additional numerical factor $\approx1.4$ arising from the finite contribution of $\lambda_2$ to $S(\omega)$), and can therefore become arbitrarily small for weak pumping.  Moreover, the emission remains centered at zero detuning, whereas in the symmetric case the peak position shifts with $N$, $w$, and $\Gamma$, cf. panel~(f).

Figure~\ref{fig:phase_plot}(h) summarizes the different regimes of model~\eqref{eqref:model} in terms of photon statistics and linewidth. By tuning $\phi$ and $w$, the emitted light can be bunched ($g^{(2)}(0)>1$) with an ultranarrow linewidth $\Delta\nu<2\Gamma$ (B UN), bunched with a narrow linewidth $\Delta\nu<\Gamma N$ (B N), bunched with a broad linewidth $\Delta\nu>\Gamma N$ (B B), or quantum ($g^{(2)}(0)<1$) with ultranarrow (Q UN), narrow (Q N), or broad (Q B) linewidth. Coherent light with $g^{(2)}(0)=1$ is also possible, but it occurs only in a fine-tuned region at the boundary between the quantum and bunched regimes shown in panel (h).

In Appendix~\ref{sec:pump_J}, we also show the result for the system in which, instead of pumping the spin $\sigma$, the incoherent pump is applied to the spin $J$. We show that tuning the relative phase in such a system can be used to generate ultranarrow-linewidth light with thermal-like second order coherence. In this case, there is not much control over the statistics of the light, but the linewidth can be tuned arbitrarily. 

\subsection{Connection to single-emitter experiments\label{sec:single_emitter}}

The two-spin case of  Sec.~\ref{sec:2spins} admits a direct experimental analog in semiconductor quantum-dot (QD) systems, where multiple emission channels coexist and can be addressed independently.
In experiments combining resonant excitation with weak above-band illumination of a single QD~\cite{Solomon,Gazzano:18}, the above-band laser populates the emitter via phonon-assisted relaxation, providing an incoherent excitation channel that coexists with the resonantly driven emission.
The above-band channel has been shown to modify both the second-order coherence~\cite{Solomon} and the emission linewidth~\cite{Gazzano:18}, qualitatively consistent with the interference mechanism analyzed above.
In the language of \eqref{eqref:model}, the resonantly-coupled and above-band-populated pathways play the roles of the unpumped spin $J$ and the pumped spin $\sigma$ respectively, with the radiative reservoir providing the shared decay channel.
A more literal realization of the $J = 1/2$ sector could be obtained in two-QD cavity-QED platforms, with one emitter pumped above-band and the cavity enforcing collective decay.

Within these platforms, the relative phase $\phi$ can be implemented in detection via polarization optics when the two channels couple to orthogonal dipoles, as demonstrated for orthogonal-dipole transitions in single QDs~\cite{Solomon2} and originally proposed in atomic systems~\cite{von_Zanthier_optica}.
Our model then makes several predictions accessible to current QD experiments:
(i) for $\phi \to \pi$ the linewidth scales as $\Delta\nu \propto \min(w, N \Gamma)$ with the incoherent pump rate $w$, narrowing toward $w$ at weak pumping;
(ii) in the symmetric channel the second-order coherence varies as $g^{(2)}(0) \approx 1.7\,N \Gamma/w$, crossing unity near $w \approx 1.7\,N \Gamma$; the $g^{(2)}(0) > 1$ regime here reflects rare two-photon emission bursts on a strongly suppressed single-photon background, which is distinct from thermal statistics, as confirmed by the behavior of higher-order coherences below;
(iii) for $\phi \approx 0$ the spectral function acquires a dip at zero detuning that fills in continuously as $\phi \to \pi$.
Most distinctively, in the weakly pumped bunched regime our model predicts that the higher-order coherences $g^{(k)}(0)$ remain dominated by two-photon processes and do not grow rapidly with $k$, in contrast to the $k!$ scaling characteristic of thermal light.
This prediction falls directly within the reach of photon-number-resolving detection schemes recently demonstrated for higher-order photon-state classification in QD emission~\cite{Thomay_multiphoton}, offering a clean experimental discriminator between the interference mechanism described here and a thermal-bath origin of the bunching.

\bigskip

The results discussed in this section highlight the importance of controlling the relative phase when measuring the emitted light in the dissipation-only system. In the analysis above, the phase $\phi$ was imprinted by hand (a possible implementation can be found in the Appendix~\ref{sec:exp}). In realistic situations, however, such a phase can also arise naturally from coherent interaction between atoms. The interaction sets a reference phase and provides an additional mechanism for controlling observable properties of the emitted light. The next section is therefore devoted to exploring the role of coherent interaction in the partially pumped system.

\section{The role of coherent interactions\label{sec:int}}

We now fix the relative phase to $\phi=0$ and include a coherent interaction. We add a coherent channel which allows an additional exchange of excitations between pumped and unpumped spins, and is
described by the Hamiltonian
\begin{equation}
H=V(J^+\sigma^-+\sigma^+ J^- ).
\end{equation}
The dynamics of the density matrix are then given by
\begin{equation}\label{eq:int}
\dot\rho= -\mathrm{i}[H,\rho]+\frac{\Gamma}{2}\mathcal{D}\left[\sigma^-+J^-\right]\,\rho
+\frac{w}{2}\mathcal{D}\left[\sigma^+\right]\,\rho.
\end{equation}
{As we show in this section, the coherent interaction leads to two main effects. First, it enables new phases featuring population inversion, $\langle S^z\rangle \ge 0$, in contrast to the toy model~\eqref{eqref:model}, where the magnetization is always negative. Second, although there is no one-to-one correspondence between $\phi$ and the ratio $V/\Gamma$, the coherent interaction has an effect qualitatively similar to that of the phase in Eq.~\eqref{eqref:model}. At the mathematical level, it renders the coherences of the steady state density matrix $\rho_{ss}$ complex, leading to interference effects analogous to those observed in the toy model~\eqref{eqref:model}. In this way, tuning the interaction strength allows the system to access different spectral and photon-statistics regimes. Below, we discuss these two effects in more detail, first by analyzing the origin of population inversion and then by summarizing the dynamical regimes accessible in model~\eqref{eq:int} as the interaction strength $V$ and pump rate $w$ are varied.}

\begin{figure*}
    \centering
    \includegraphics[width=1\linewidth]{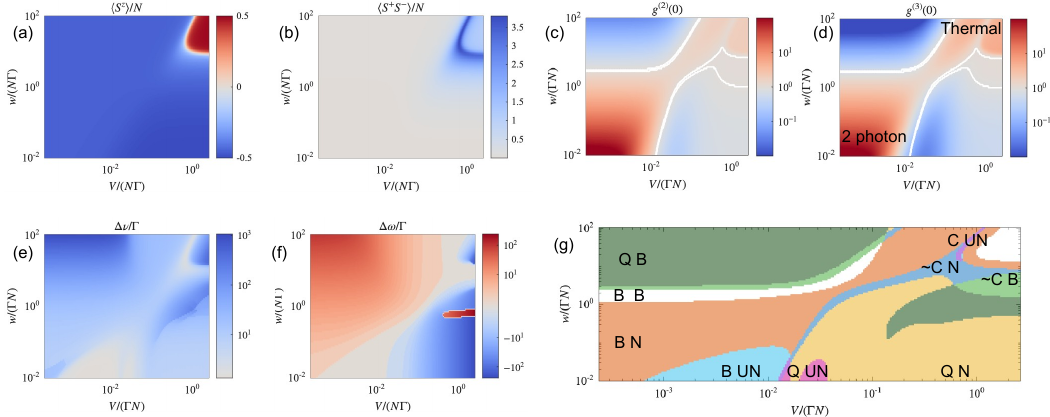}
    \caption{ \textbf{Properties of the light emitted by an ensemble of  $21$ atoms in the presence of coherent interaction $V$.} 
    Steady-state (a) magnetization, (b) intensity, (c) second- and (d) third-order coherence, (e) linewidth, and (f) position of the peak in the spectral function as functions of the pump rate $w$ and interaction strength $V$. (g) Diagram summarizing the regimes with bunched light $g^{(2)}(0)>1$ and ultranarrow 
     $\Delta\nu<2\Gamma$ 
   (B UN), narrow  $\Delta\nu<\Gamma N$ (B N), broad $\Delta\nu>\Gamma N$ (B B) linewidth,  classical (or coherent) light $g^{(2)}(0)\to1$ with ultranarrow  (C UN, superradiant lasing regime),  narrow  (C N) and broad (C B) linewidth and quantum light $g^{(2)}(0)<1$ 
   with ultranarrow (Q UN), narrow (Q N) and broad (Q B) linewidth.
     }
    \label{fig:int}
\end{figure*}

\subsection{Role of interaction in inducing population inversion}
If the pump rate is within the limits $\Gamma N^2/2 \lesssim w\lesssim 4V^2/\Gamma$, the steady state of the model~\eqref{eq:int} enters the regime with positive magnetization, $\langle S^z\rangle\ge 0$.
The lower threshold $w$ value is set by the condition that the pump must be strong enough to compete with the collective loss, which yields the scaling $w\gtrsim \Gamma N^2/2$ {(notice the $\propto N^2$ scaling of the loss term in the Lindblad master equation, compared to the pump term, which acts only on a single spin)}. {Another way to see the origin for  $w\gtrsim \Gamma N^2/2$ condition is the following: To achieve population inversion, the interaction must be strong enough to transfer approximately $N/2$ excitations to the spin $J$ before collective emission occurs, $V\gtrsim \Gamma N/2$. At the same time, the incoherent pump must be sufficiently strong to repopulate the spin $\sigma$ after it transfers excitation to $J$, $w\gtrsim N V/2$; otherwise, the inverse process $\propto V \sigma^+ J^-$ is activated, keeping the total number of excitations approximately constant. It sets the following hierarchy for energy scales $w\gtrsim NV\gtrsim N(\Gamma N)$.

The upper bound on the pump rate can be derived by adiabatic elimination of the spin $\sigma$ (valid when $w\gg V,\,\Gamma$)~\cite{eff_opp}, which yields the effective master equation for the collective spin
\begin{equation}\label{eq: effective model}
\dot \rho \approx \frac {\Gamma}{2}\mathcal D[J^-]\,\rho +\frac{w}{2}\frac{(2V)^2}{w^2}\mathcal {D}[J^+]\,\rho+\order{\frac{\Gamma V}{w}},
\end{equation}
for which population inversion appears when the effective pump rate exceeds the loss rate, which yields the condition on the upper boundary for the pump rate, $w<4V^2/\Gamma$. 
We observe that the effective pump rate for the $J$ spin in Eq.~\eqref{eq: effective model} decreases with increasing $w$. Indeed, in the limit when the pump sets the largest energy-scale, the pump effectively projects the spin $\sigma$ onto the $\ket{\uparrow}$ state, suppressing the off-diagonal components of the steady-state density matrix. This projection suppresses the expectation values $\langle \sigma^{\pm}\rangle$ and effectively tends to decouple the spins $\sigma$ and $J$. Thus, when the pump rate exceeds the critical value $4V^2/\Gamma$, the coherent evolution in Eq.~\eqref{eq:int} becomes too ineffective to compensate the losses, and population inversion cannot be obtained. This regime can be viewed as an analogue of the Zeno effect~\cite{breuer2002theory}: the strong pump continuously projects the spin $\sigma$ onto the $\ket{\uparrow}$ state and effectively freezes the coherent dynamics.

\subsection{Summary of dynamical regimes }

In Fig.~\ref{fig:int} we plot steady-state observables of the interacting model for $N=20$ spins as functions of the interaction strength $V$ and pump rate $w$. For $V=0$, the data reproduces the  $\phi=0$ case of the non-interacting model. The magnetization in panel (a)  features  population inversion  in a wedge-shaped region of the parameter space, as discussed above. 
The intensity of the emitted light in panel (b) increases with both interaction strength and pump rate, with maxima corresponding to the region of near-zero magnetization in panel (a). There, the intensity scales as $\propto N^2$, similarly to the case of superradiant emission. As the pump rate increases further, the intensity decreases to $\propto N$ when all spins become inverted.
 
The second and third order coherence  in panels (c,d) show regions corresponding to all three photon-statistics regimes, depending on $w$ and $V$. At weak pump, the light evolves from bunched to quantum as $V$ increases, similarly to the effect of varying $\phi$.
In addition, for weak pump the bunched regime is dominated by two-body correlations, while at strong pump higher-order correlations become important, reminiscent of thermal light statistics. For stronger pump and interaction strength, panel (c) also shows the regions with $g^{(2)}(0)\to 1$, where $V$ is sufficient to keep emission coherent.

Panels (e) and (f) show the linewidth $\Delta\nu$ and the frequency shift $\Delta\omega$ of the main peak in the spectral function of the emitted light respectively. Overall, interaction modifies the shape of the spectral function significantly, compared to the dissipation-only case. Some examples are presented in Appendix~\ref{sec:S_omega_int}. The similarity persists only in $V\ll w$ region, when the dip at zero detuning survives. As the interaction increases, the corrections to the shape of spectral function become non-perturbative. First, the interaction naturally induces line shifts for different eigenmodes of $\mathcal{L}$, as they acquire imaginary part. Second, as $V$ increases, the total number of excitations in the system also grows, allowing for interference between different modes which results in multiple peaks in the shape of $S(\omega)$. 
 Interestingly, the ultranarrow linewidth appears not only in the weak-pump regime $w\propto\Gamma$, but also in a region around $w\approx\Gamma N^2$ (where magnetization approaches zero 
and $g^{2}(0)\to 1$), where the linewidth becomes suppressed even for strong pumping. This regime is reminiscent of a superradiant lasing  phase~\cite{Holland_SR_laser}, even though only a single spin is incoherently pumped. In this case, coherent interactions are essential, as they enable the emergence of this phase.

Summarizing the results in Fig.~\ref{fig:int}(a–f), we can distinguish several regimes, see panel (g). Importantly, the interacting model hosts all phases of our toy model even though here the ``phase'' $\phi$ is induced only by the interaction $V$.  
For zero and near-zero interaction strength, the phase diagram recreates regimes of the toy model in the $\phi\to  0 $ limit, and finite interaction strength  allows additionally to reach bunched ultranarrow, quantum ultranarrow and quantum narrow regimes, analogous to the effect we see when varying $\phi\to \pi $ in the toy model \eqref{eqref:model} (although the ultranarrow linewidth can be observed only in a more restricted parameter window).
Additionally, interaction stabilizes regimes where system emits coherent light, indicated as classical ultranarrow (C UN), narrow (C N) and broad (C B) regions in Fig.~\ref{fig:int}(g).
The rich variety of accessible regimes suggests that the interplay between inhomogeneous incoherent pumping, collective loss and coherent interaction provides a powerful mechanism for controlling the properties of light sources.

\section{Discussion and Outlook\label{sec:discussion}}

Summarizing our results, we have shown that breaking permutation invariance through spatially inhomogeneous pumping allows the system to access dynamical regimes that are absent in fully symmetric settings. In permutation-invariant systems, light emission is effectively governed by a single collective channel, whereas symmetry breaking creates different emission pathways and thereby enables interference. In this regime, the transfer of excitations from the pump to the environment proceeds through multiple channels, and the interference between them becomes the key mechanism determining the spectral and statistical properties of the emitted light. Crucially, introducing a relative phase between these channels allows one to continuously tune the interference from destructive to constructive, leading to strong modifications of observables and providing a simple route to engineer desired emission properties in a single platform.

As we have shown in this work, even in a minimal model where only one spin is pumped (or, equivalently, one spin is left unpumped), control over the phase and pump rate already gives access to qualitatively distinct regimes of light emission. Depending on parameters, one can generate quantum ultranarrow light (Sec.~\ref{sec:dissip_only}), coherent ultranarrow light (Sec.~\ref{sec:int}), or thermal ultranarrow light (Appendix~\ref{sec:pump_J}), whereas in the absence of such control the emitted light remains broad or has completely different properties. These results illustrate that partially pumped systems provide a versatile platform in which narrow-linewidth emission can emerge from mechanisms beyond conventional symmetric lasing scenarios. 
Beyond the many-body setting, the simplest signatures of this interference mechanism, the predicted scaling of $g^{(2)}(0)$ with the incoherent pump rate, and the higher-order coherence structure that distinguishes the bunching from thermal emission [see Sec.~\ref{sec:single_emitter}] are accessible in current single-emitter and few-emitter cavity-QED experiments using photon-number-resolving detection~\cite{Thomay_multiphoton}, offering a near-term route to quantitative validation.

More broadly, our findings demonstrate that dissipation need not merely act as a source of decay and dephasing, but can instead serve as a resource for generating coherence and correlations. In the models~\eqref{eqref:model} and~\eqref{eq:int}, coherence is dynamically maintained through the interplay of collective loss and local pumping. The collective loss channel imprints correlations between emitters, while the coexistence of several emission paths gives rise to interference effects that shape measurable observables. This perspective is consistent with a growing body of work showing that dissipative dynamics can stabilize new steady states, generate unconventional many-body phases, or even emulate paradigmatic condensed-matter phenomena~\cite{Marino_universality,Kondo1,Kondo2,PhysRevB.108.104302,dey2026dissipationmechanismsdissipativephase}. Correlations generated by collective dissipation may also have practical applications, as illustrated by recent proposals employing boundary time crystals for AC sensing~\cite{Iemini_AC}.

An important aspect of our setup is that the inhomogeneous drive explicitly breaks permutation invariance by separating the system into pumped and unpumped subsystems. The role of symmetry breaking in many-body systems has attracted increasing attention. Even when introduced only through initial conditions, non-permutation-invariant sectors may support dynamics qualitatively different from those of the symmetric manifold, including persistent coherences or signatures of quantum chaos~\cite{Ziolkowska2025deephilbertspacealltoall,Iemini_Chang_Marino}. In our case, the symmetry breaking is built directly into the dissipative dynamics, so that even an initially permutation-invariant state can evolve within a richer effective Hilbert space and display interference phenomena unavailable in fully symmetric models.

Interestingly, recent studies of lasing in partially pumped systems indicate that coherent interaction may play contrasting roles depending on the microscopic implementation. In Ref.~\cite{Bruder_laser}, the authors studied emission from atoms in a cavity where only $N_p$ out of a total of $N$ atoms are pumped. There, interaction was found to be detrimental, leading to linewidth broadening and even suppression of the lasing phase. A possible explanation is the hierarchy of energy scales required for realizing lasing in such partially pumped system, since the minimal pump rate for lasing regime scales as $\Gamma N^2/N_p$, implying a substantially stronger incoherent drive than in the homogeneously pumped case, while the interaction should exceed $\Gamma N$, to ensure transfer of excitations to unpumped emitters. 
The precise form of the interaction is also crucial. For example, for an interaction term of the form $V(J^+ + \sigma^+)(J^-+\sigma^-)$, which naturally arises after elimination of a leaky cavity mode,  no linewidth narrowing in lasing phase is observed in our toy model. On the other hand, in the absence of the $J^+J^-$ contribution, ultranarrow light can be emitted even when only a single atom is pumped. 
By contrast, Ref.~\cite{Ritsch_partial_laser} showed that in free space, partial pumping together with coherent interaction can enable mirrorless lasing. There, dipolar interaction between atoms allow excitations to be transferred to the unpumped emitters, while pump-induced noise affects only the pumped atoms, resulting in a narrower linewidth than in the homogeneously pumped case. 
These contrasting observations in Refs.~\cite{Bruder_laser,Ritsch_partial_laser} highlight the importance of identifying the microscopic mechanisms that govern emission in spatially inhomogeneous systems, including the interplay of interaction, noise, and interference.

Our results also suggest several promising future directions. Modern experimental platforms such as arrays of Rydberg atoms, solid-state quantum-optical interfaces, and optical tweezer arrays offer the possibility to realize more complex geometries in which dissipation has a finite correlation length, potentially leading to even richer interference effects~\cite{PhysRevB.111.064424,douglas2026manybodysupersubradianceordered,Ritsch_partial_laser}. Correlated incoherent processes may reveal broad classes of emission regimes inaccessible in single-channel systems, while related interference mechanisms could be explored in multilevel emitters as well~\cite{Solomon2}. We therefore expect partially pumped driven-dissipative systems to provide a fruitful setting for engineering unconventional light sources and uncovering new forms of non-equilibrium many-body dynamics.

\begin{acknowledgments}
We thank Helmut Ritsch for fruitful discussions and valuable suggestions.  JM acknowledges support from the CAS Dean’s office at SUNY Buffalo. This study was financed, in part, by the São Paulo Research Foundation (FAPESP), Brasil, Process Number 2024/22542-4.
We acknowledge generative AI for assistance with coding and with manuscript preparation. 
\end{acknowledgments}

\appendix

\section{Exact and semiclassical solutions\label{sec:ED}}

The model~\eqref{eqref:model} can be solved efficiently by initializing the system in a state belonging to 
 the symmetric spin $J$ manifold. The model~\eqref{eqref:model} conserves the total angular momentum $J$, and thus the basis for the collective spin contains $N+1$ Dicke states, for which $J^{\pm}|m\rangle=\sqrt{J(J+1)-m(m\pm 1)}|m\pm 1\rangle$, $J^z|m\rangle =m|m\rangle$, $m=-J,\ldots, J$. The pumped spin $\sigma$ has a basis consisting of two states, $\ket{\uparrow}$ and $\ket{\downarrow}$. As such, the density matrix can be expressed as 
\begin{equation}
    \rho=\sum_{m,m'=-J}^{J}\sum_{s,s'=\uparrow,\downarrow} \alpha_{m,m'}^{s,s'}\ketbra*{m,s}{m',s'}
\end{equation}
with $4(N+1)^2$ elements in total. Restricting the Hilbert space to the symmetric subspace for $J$ in this case significantly reduces its dimension (from $4^{N+1}$) and allows for an exact solution for larger system sizes. 

The accessible system sizes can be further increased by noting that $J^z$ remains close to $-J/2$ in the steady state (the south pole of the Bloch sphere), meaning that the dissipative coupling is not sufficient to significantly deviate it from the fully de-excited state. Therefore, we can perform a Holstein--Primakoff transformation and even limit the expansion to the lowest order~\cite{altland2010condensed}. Performing the expansion around the collective spin-down state, we can approximate
\begin{equation}
    \begin{aligned}
        J^+&\approx \sqrt{N}a^{\dagger}\\
        J^z&\approx -\frac{N}{2}+a^{\dagger}a
    \end{aligned}
\end{equation}
Now, the problem reduces to that of one spin coupled to a single bosonic mode. Because spin $J$ is not pumped directly, the solution converges to the exact one for the spin system already for a small cutoff number of bosons, $n_{cut}\ge 2$. However, when calculating higher-order coherences, a larger cutoff should be used. 

The Holstein--Primakoff transformation allows us to obtain an exact numerical solution of the model for rather large system sizes with low computational cost.  In this way, we can ensure that the results reported in this paper are not finite size effects. Note, that for the model with coherent interaction~\eqref{eq:int}, the Holstein-Primakoff approximation becomes invalid rapidly with $V$ increased and the exact solution in the spin basis is required.

\section{Connection to dissipative Jaynes–Cummings model and semiclassical solution\label{sec:semiclassics}} 
After bosonizing the collective spin using the Holstein–Primakoff approximation, the master equation becomes a version of dissipative  Jaynes–Cummings  model~\cite{raimond2006exploring}
\begin{equation}\label{eq:HP_model}
\mathcal L\tilde\rho=
\frac{\Gamma}{2}\mathcal D\left[e^{i\phi}\sigma^-+\sqrt{2J} a\right]\,\tilde\rho
+\frac{w}{2}\mathcal D[\sigma^+]\,\tilde\rho,
\end{equation}
and describes a single incoherently pumped spin dissipatively coupled to a single bosonic mode.

Using a cumulant expansion~\cite{kerber2025cumulantsexpansionapproachgood}, one can recover the scaling of the relevant Liouvillian modes with $w$, $\Gamma$, and $N$, which explains the crossover between narrow and broad spectral features discussed below. In the limit $w\gg\Gamma$, where the magnetization of spin $\sigma$ can be approximated by $\langle \sigma^z\rangle\to 1/2$, the two leading eigenmodes are
\begin{equation}
\begin{aligned}
\lim_{w\gg \Gamma }\lambda_{1,2}&\approx -
\frac{1}{4}\Big(w+\Gamma (N+1) \\
&\mp \sqrt{-4N \Gamma (w+2\Gamma )+(w+\Gamma (N+1))^2}\Big)
\end{aligned}
\end{equation}
Here, the slowest eigenvalue tends to $\Gamma(N+1)/2+1/w(\ldots)$, while the second-slowest eigenvalue scales as $\propto (w+\Gamma (N+1))/2$ in the large-$w$ limit, in agreement with exact numerics.

Although the cumulant expansion does not capture all features of the problem, such as exceptional points, it provides a good estimate of the shape of the spectral function and other observables. Alternatively, the scaling properties can be extracted from the exact numerical simulations.

\begin{figure}
    \centering
    \includegraphics[width=1\linewidth]{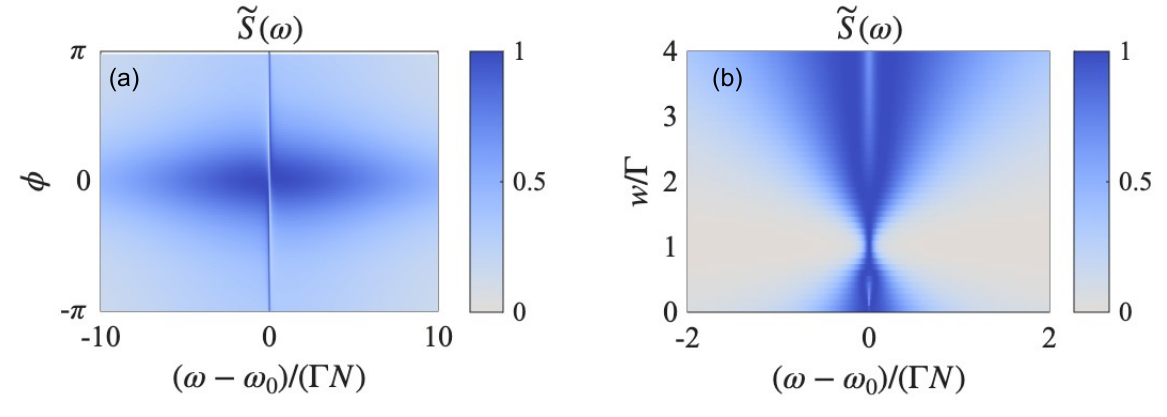}
    \caption{ 
    Normalized spectral function for system of $21$ spins governed by model~\eqref{eq:coll_model} as a function of (a) $\phi$ for $w=\Gamma N$ and (b) as a function of pump rate for $\phi=0$.}
    \label{fig:coll_sp_f}
\end{figure}

\begin{figure*}
    \centering
    \includegraphics[width=0.95\linewidth]{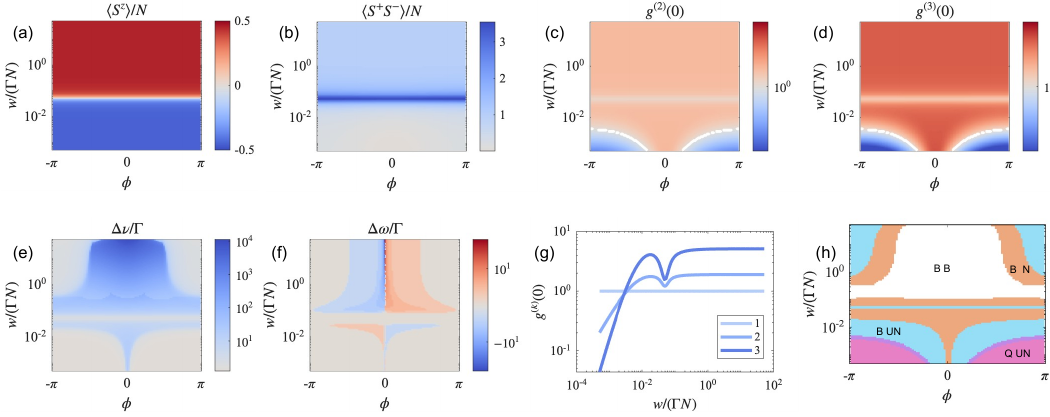}
    \caption{\textbf{Properties of the light emitted by an ensemble of $21$ atoms governed by model~\eqref{eq:coll_model}.} 
    Steady-state (a) magnetization, (b) intensity, (c) second-order coherence, (d) third-order coherence, (e) linewidth, and (f) position of the peak in the spectral function as functions of the pump rate $w$ and the phase $\phi$. (g) Dependence of the $k$-th order coherence on the pump rate for different orders $k$ and $\phi=\pi$. The light displays thermal statistics in the bunched regime. 
    (h) Diagram summarizing the regimes with bunched light $g^{(2)}(0)>1$ and ultranarrow linewidth $\Delta\nu<2\Gamma$ (B UN), bunched light with narrow linewidth $\Delta\nu<\Gamma N$ (B N), bunched light with broad linewidth $\Delta\nu>\Gamma N$ (B B), and quantum (antibunched) light $g^{(2)}(0)<1$ with ultranarrow (Q UN) linewidth.}
    \label{fig:coll_plot}
\end{figure*}
 
\section{Toy model for generating thermal ultranarrow light\label{sec:pump_J}}

In this section, we briefly consider another solvable limit of partially pumped systems, in which the incoherent pump is applied to the collective spin $J$, rather than to a single spin. As we show, this case also exhibits linewidth narrowing in the limit $\phi \to \pi$; however, the second-order coherence approaches $g^{(2)}(0) \to 2$ over most of the parameter range, corresponding to the photon statistics of thermal light.

The dynamics of the partially pumped system are governed by the following master equation
\begin{equation}\label{eq:coll_model}
    \dot\rho=\frac{\Gamma}{2}\mathcal{D}[e^{\mathrm{i}\phi}J^-+\sigma^-]\,\rho+\frac{w}{2}\mathcal{D}[J^+]\,\rho
\end{equation}
Without the spin $\sigma$, the model with collective pump and loss of spin $J$ can be solved exactly in the basis of the symmetric Dicke manifold $|J,m\rangle$. The solution is a mixture of Dicke states with fixed $J=N/2$, $\rho_{ss}=\sum_{m}\alpha_{m}|m\rangle\langle m|$.  The coefficients $\alpha_m$ can be derived as $\alpha_m=r^{m}r^{J}\alpha_{-J}$, where $r=w/\Gamma$. The normalization then imposes the constraint $\sum_m\alpha_m=1$. Interestingly, the system satisfies detailed balance~\cite{detailed_balance}, and the steady state $\rho_{ss}$ is identical to a thermal state for a system with Gibbs distribution $\rho=\exp(-\beta H)/Z,$ with $H=\omega_0 J^z$. Here, $\alpha_m=\exp(-\beta\omega_0 m)/Z=r^{m+J}\alpha_{-J}$, see also Ref.~\cite{shimshi2024dissipativephasetransitionmetrology}. The distribution of $\alpha_m$ in both cases is the same if one introduces the effective temperature $\beta\omega_0=\ln(\Gamma/w)$. Specifically, it corresponds to a thermal distribution with positive temperature for $\Gamma>w$, negative temperature in the population-inverted regime $w>\Gamma$, and infinite temperature for $w=\Gamma$. For finite $T$, the second-order coherence $g^{(2)}(0)\to 2$ corresponds to that of thermal light. The minimal possible value of the second order coherence is $6/5$ in the $T=\infty$ ($\beta=0$) case~\footnote{{We note that this effective temperature describes a state confined within the Dicke manifold at a fixed $J$, so it still has much more structure than the maximally mixed state extended to the whole Hilbert space.}}. In this regime, $\langle J^+ J^-\rangle \propto N^2$, and the linewidth of the emitted light is $\Gamma+w=2\Gamma$. For finite temperature, instead, $\langle J^+ J^-\rangle$ scales as $N$ for $w>\Gamma$ or as $N^0$ for $w\ll\Gamma$, and the linewidth grows as $\Delta\nu=N|w-\Gamma|$.

When we include the additional spin $\sigma$, this allows for further control over observables, especially if we can tune the relative phase $\phi$ between $J$ and $\sigma$ in the dissipation channel (or, equivalently, in the observables $S^{\pm}$). Figure~\ref{fig:coll_sp_f}(a) shows the normalized spectral function as a function of $\phi$ for $w=\Gamma N$. Here, there is a significant difference in the linewidth between the $\phi=0$ case, $\Delta\nu \propto |w-\Gamma|N$, and the $\phi\to\pi$ case, $\Delta\nu\propto \Gamma$.
Interestingly, when $w\approx \Gamma$, the dip in the spectral function vanishes even for $\phi=0$, see Fig.~\ref{fig:coll_sp_f}(b). We explain this effect by noting that in this limit the system is close to a maximally mixed state  and thus has a finite population of states with larger number of excitations. In this regime, the coherences are suppressed and bring only minor corrections to the total shape of $S(\omega).$ 
As the pump rate increases beyond $\Gamma$, the number of excitations in the system become smaller again and the relative contribution from the  singlet-like single excitation sector  to the $S(\omega)$ increases resulting at the dip reappearing at larger pump rates.

In Fig.~\ref{fig:coll_plot} we summarize observables as functions of $\phi$ and $w$ for a system with $N=20$. Here, we limit the system size in order to explore the region around $w=\Gamma$, where the Holstein--Primakoff approximation breaks down. However, near-exact numerics can be performed for arbitrary system sizes deep in the $w\ll\Gamma$ and $w\gg \Gamma$ regimes. 

The magnetization $\langle S^z\rangle$ in panel (a) displays a rapid transition from $-J-1/2$ to $J+1/2$ around $w=\Gamma$, as in the collective model. The intensity in panel (b) grows from $\propto N^0$ when $w\ll\Gamma$ to a maximum $\propto N^2$ at $w=\Gamma$, and then scales as $\propto N$ for $w>\Gamma$. The emitted light mostly exhibits statistics characteristic of thermal light emitted by an ensemble of atoms. In panel (c), $g^{(2)}(0)\to 2$ for most parameter values, with a sharp dip to $g^{(2)}(0)\approx 6/5$ when $w\to\Gamma$. There is also a small region of quantum light when $\phi\to\pi$ at weak pump. The light statistics are confirmed by evaluating $g^{(3)}(0)$ in panel (d). Here, $g^{(k)}(0)$ increases with $k$ in the bunched regime, indicating the buildup of multi-particle correlations, and decreases with $k$ in the quantum regime, indicating suppression of multiphoton processes. We plot $g^{(k)}(0)$ for different values of $k$ at $\phi=\pi$ in panel (g). As $w$ increases, the system transitions from quantum light to a bunched regime with reduced values of $g^{(k)}(0)$ around $w=\Gamma$, and then to a regime where $g^{(k)}(0)$ increases and saturates to values corresponding to thermal light.

Panel (e) shows the dependence of the linewidth $\Delta\nu$ on $\phi$ and $w$. In the $\phi\to 0$ case, the linewidth follows the same scaling as in the collective model, starting from $(\Gamma-w)N$, reaching a minimum slightly above $2\Gamma$ at $w=\Gamma$, and then increasing again as $(w-\Gamma)N$. For stronger pumping, when two peaks appear in the spectral function, the FWHM changes to $(w-\Gamma)N/2$. In the  $\phi\to\pi$ case, the linewidth starts at $\Delta\nu \approx 1.2\Gamma$ for weak pump,  then grows linearly with the pump rate up to the region $w\propto \Gamma$, where it decreases back to $2\Gamma$. Between $w=\Gamma$ and $w=\Gamma N$, the linewidth decreases further, approaching $\Gamma$ as $w\to\infty$.  The spectral function  exhibits a line shift for $\phi\to 0$ due to the double peak structure, while in $\phi\to\pi$ region the spectral function is centered at zero detuning, see panel (f). Note that, in the vicinity of $w=\Gamma$ region, the spectral function contains a single peak for any $\phi$, so the line shift is zero.

The observation that the real part of the slowest eigenmode does not scale with $N$ and results into $\Delta\nu\to\Gamma$ for $\phi\to\pi$  has a simple explanation. As discussed in Sec.~\ref{sec:dissip_only}, the slowest mode is associated with the unpumped subsystem, which is subject only to noise from the loss process. Here, this subsystem is a single spin $\sigma$ with decay rate $\Gamma$, in contrast to the collective spin $J$ in Sec.~\ref{sec:dissip_only}, which has an effective decay rate $\Gamma N$. As a result, the linewidth remains intensive even at strong pump in the $\phi = \pi$ case.

Panel (h) summarizes the operational regimes of the model~\eqref{eq:coll_model}, highlighting changes in light statistics as well as control over the linewidth. 
The model hosts a small region of quantum ultranarrow light, as well as bunched light with ultranarrow, narrow and broad linewidth. 
The most interesting regime takes place in the vicinity of $\phi\to\pi$ region, which  shows an extended region of  thermal-like light with an ultranarrow linewidth (B UN). Interestingly, decoupling just one spin allows the stabilization of a region with ultranarrow linewidth, compared to the fully collective model, where $\Delta\nu=2\Gamma$ occurs only at a single fine-tuned point $w=\Gamma$.

\section{A proof-of-principle implementation route\label{sec:exp}}

The central requirement of the model is the ability to control the relative phase entering the dissipative channel or the measured output field. In most direct physical realizations, however, the phase imprinted in the collective jump operator is automatically inherited by the emitted field, which makes independent control of these phases difficult. As a result, the interference effects with $\phi\ne 0$ discussed in the main text are not straightforwardly accessible in a single-channel setup.

A possible route to overcome this limitation is to introduce additional weak auxiliary dissipation channels that serve primarily as readout ports. Concretely, one may consider a cavity-QED setup~\cite{mivehvar2021cavity} in which the atoms are coupled to a primary cavity mode that realizes the collective decay channel
\[
\frac{\Gamma}{2}\,\mathcal{D}[J^-+\sigma^-]\,\rho,
\]
while two additional auxiliary modes (can be from auxiliary cavities) couple separately to $\sigma^-$ and $J^-$. These auxiliary channels generate extra weak dissipation terms of the form
\[
\frac{\kappa}{2}\,\mathcal{D}[\sigma^-]\,\rho+\frac{\kappa}{2}\,\mathcal{D}[J^-]\,\rho,
\]
with $\kappa \ll \Gamma$. The output fields of the auxiliary modes can then be interferometrically combined with a controllable relative phase, effectively engineering the measured operator without significantly modifying the dominant Liouvillian dynamics.

In this implementation, the primary collective cavity defines the dissipative evolution of the system, while the auxiliary channels provide access to the emitted fields for post-processing. By introducing a phase shift between the auxiliary output channels, one can construct a detection operator of the form
\[
S^-= J^- + e^{\mathrm{i}\phi}\sigma^-,
\]
thus reproducing the phase-dependent interference effects discussed in the main text at the level of measured observables.

We have verified numerically that the main qualitative features of the model persist in the presence of such auxiliary channels, provided that $\kappa$ remains sufficiently small compared to $\Gamma$. In particular, the interference-induced modifications of the spectral function and photon statistics remain visible for $\kappa \lesssim \Gamma/10$, and partially survive even for $\kappa \sim \Gamma/2$, although in the latter case stronger pumping is required to compensate for the additional local losses.
The above scheme assumes that the different cavity modes can be treated as independent and that cross-couplings between them are negligible. We also assume that the auxiliary channels are sufficiently weak so that they do not significantly perturb the Liouvillian spectrum, while still providing measurable output fields for interferometric detection.
While this setup does not provide independent control over the phase in the dissipative channel itself, it offers a realistic way to access the interference effects predicted by the model through phase-sensitive detection. In any case, the realistic implementation will contain the coherent interaction as well as the dissipation, which reduces the necessity to control phase $\phi$ at all.  However, auxiliary cavities may still be in use to suppress the $J^+J^-$ coherent interaction term.

\begin{figure}
    \centering
    \includegraphics[width=1\linewidth]{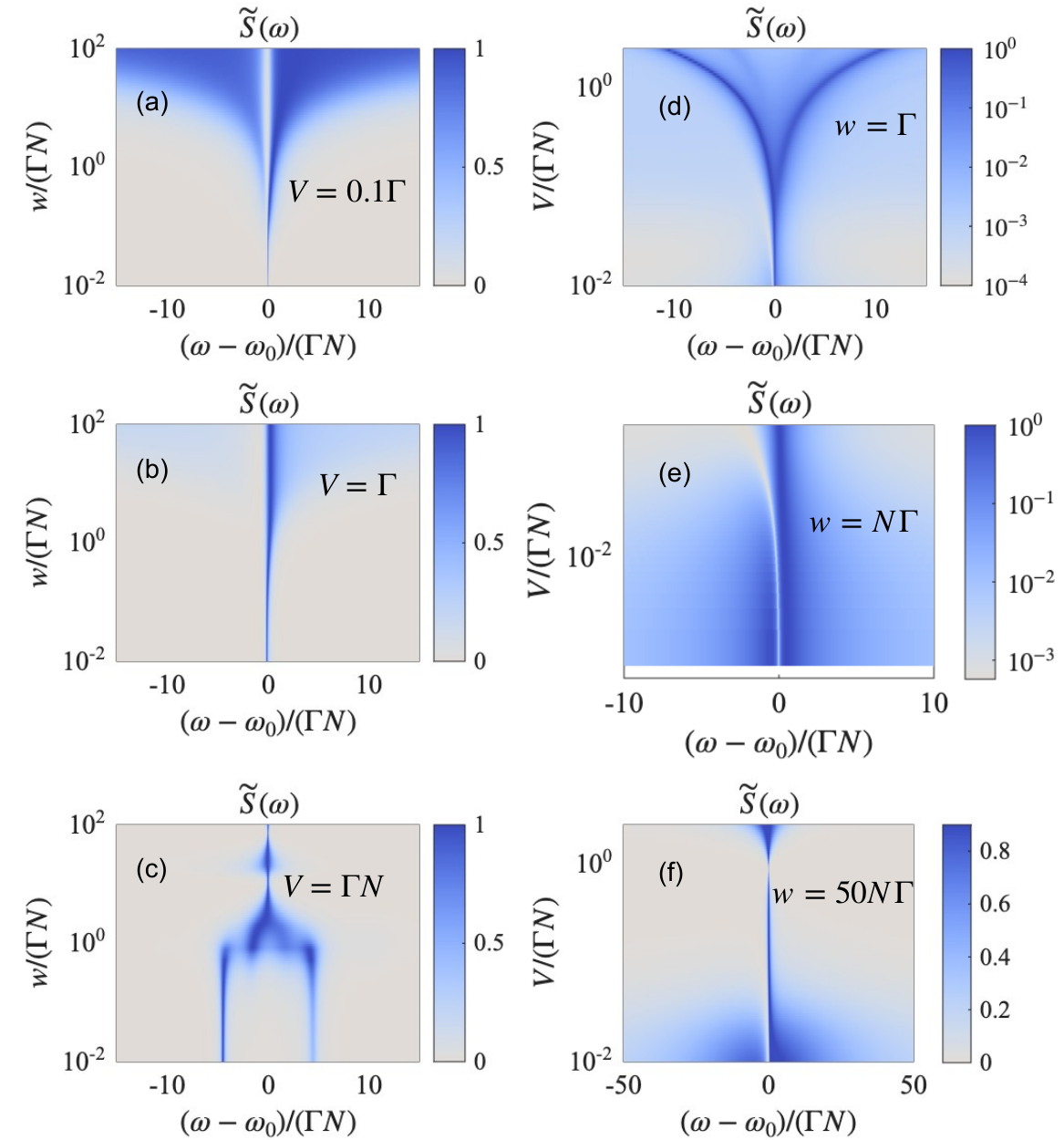}
\caption{ \textbf{Normalized spectral functions for the partially pumped system with coherent interaction.}
Normalized spectral functions for model~\eqref{eq:int} as a function of pump rate (a–c) and interaction strength (d–f).
Notice, that $\tilde S(\omega)$ in panels (d) and (e) is plotted on a logarithmic scale to emphasize fine details.}
\label{fig:interacting_S_omega}
\end{figure}

\section{Spectral functions for the model with coherent interactions\label{sec:S_omega_int}}

In this Appendix, we show several spectral functions for model~\eqref{eq:int} for different values of $V$ and $w$. This analysis is intended to summarize when $V$ can be treated as a small perturbative dressing of our toy model~\eqref{eqref:model}, and when the interaction introduces non-perturbative modifications to the shape of the spectral function.

First, we consider the dependence of the spectral function on the pump rate for different interaction strengths $V$, see Fig.~\ref{fig:interacting_S_omega}. For $V=0.1\Gamma$ in panel (a), the spectral function preserves a double-peak structure for $w\gg V$, while for weak pumping the interaction, and hence the emergence of complex coherences, leads to a single-peak structure. For $V=\Gamma$ in panel (b), most parameter regimes exhibit a single peak with a positive line shift; however, a double-peak structure re-emerges at strong pumping. For $V=\Gamma N$ in panel (c), the strong interaction dominates over the effect of pumping on the spectral function. In this case, for weak pump the spectral function contains two peaks corresponding to different eigenmodes (note that the interaction induces a frequency shift, so these two peaks do not originate from destructive interference, in contrast to the dissipative model). In the $w\approx \Gamma N$ limit, the spectral function  exhibits multiple peaks.

In panel~(d), we plot the normalized $S(\omega)$ as a function of interaction strength while fixing $w=\Gamma$. For $V\to 0$, the spectral function contains two peaks, but even a small finite $V$ quickly induces an imbalance in their amplitudes. For stronger interactions, additional peaks appear in the spectral function and the main peaks become frequency shifted. At higher pump rates, shown in panels (e,f), a similar behavior is observed, with the exception that the range of weak interactions over which a double-peak structure persists becomes broader.

\bibliography{lit}
\end{document}